\documentclass[dvips,12pt]{article}
\usepackage{graphicx}
\newcommand{\doublespacing}{\let\CS=\@currsize\renewcommand{\baselinesstrech}
{2.0}\tiny\CS}

\begin{document}

\textwidth 16cm
\newcommand{\bd}{\begin{document}}
\newcommand{\ed}{\end{document}}
\newcommand{\bc}{\begin{center}}
\newcommand{\ec}{\end{center}}
\newcommand{\bfr}{\begin{flushright}}
\newcommand{\efr}{\end{flushright}}
\newcommand{\lt}{\left}
\newcommand{\rt}{\right}
\newcommand{\vs}{\vspace}
\newcommand{\hs}{\hspace}
\newcommand{\beq}{\begin{equation}}
\newcommand{\eeq}{\end{equation}}
\newcommand{\lb}{\linebreak}
\newcommand{\pb}{\pagebreak}
\newcommand{\mb}{\makebox}
\newcommand{\fb}{\framebox}
\newcommand{\mc}{\multicolumn}
\newcommand{\ben}{\begin{enumerate}}
\newcommand{\een}{\end{enumerate}}
\newcommand{\bit}{\begin{itemize}}
\newcommand{\eit}{\end{itemize}}
\newcommand{\ol}{\overline}
\newcommand{\un}{\underline}
\newcommand{\lefq}{\lefteqn}
\newcommand{\ba}{\begin{array}}
\newcommand{\ea}{\end{array}}
\newcommand{\beqa}{\begin{eqnarray}}
\newcommand{\eeqa}{\end{eqnarray}}
\newcommand{\beqas}{\begin{eqnarray*}}
\newcommand{\eeqas}{\end{eqnarray*}}
\newcommand{\bfg}{\begin{figure}}
\newcommand{\efg}{\end{figure}}
\newcommand{\bds}{\begin{displaymath}}
\newcommand{\eds}{\end{displaymath}}
\newcommand{\btb}{\begin{tabbing}}
\newcommand{\etb}{\end{tabbing}}
\newcommand{\para}{\parallel}
\newcommand{\pad}{\partial}
\newcommand{\nn}{\nonumber}
\newcommand{\la}{\leftarrow}
\newcommand{\ra}{\rightarrow}
\newcommand{\lgla}{\longleftarrow}
\newcommand{\lgra}{\longrightarrow}
\newcommand{\La}{\Leftarrow}\newcommand{\Ra}{\Rightarrow}
\newcommand{\Lra}{\Leftrightarrow}
\newcommand{\Lgla}{\Longleftarrow}
\newcommand{\Lgra}{\Longrightarrow}
\newcommand{\bm}{\boldmath}
\newcommand{\lan}{\langle}
\newcommand{\ran}{\rangle}
\renewcommand{\a}{\alpha}
\renewcommand{\b}{\beta}
\newcommand{\g}{\gamma}
\newcommand{\G}{\Gamma}
\renewcommand{\d}{\delta}
\newcommand{\eps}{\epsilon}
\newcommand{\Th}{\Theta}
\newcommand{\s}{\sigma}
\newcommand{\lam}{\lambda}
\newcommand{\D}{\Delta}
\newcommand{\vare}{\varepsilon}
\newcommand{\pr}{\prime}
\newcommand{\ro}{\rho}
\newcommand{\nab}{\nabla}
\newcommand{\m}{\mu}
\newcommand{\n}{\nu}
\newcommand{\Sg}{\Sigma}
\newcommand{\p}{\pi}
\newcommand{\R}{I\!\!R}
\newcommand{\om}{\omega}
\newcommand{\Om}{\Omega}
\newcommand{\ze}{\zeta}
\newcommand{\vart}{\vartheta}
\newcommand{\tri}{\triangle}
\newcommand{\f}{\frac}
\newcommand{\iny}{\infty}
\newcommand{\pro}{\propto}
\bc {\huge \bf Generalized Swanson Models } \ec

\bc {\huge \bf and their solutions } \ec

\vs{.5cm}

\bc
{\it \large A. Sinha{\footnote {e-mail : anjana$_-$t@isical.ac.in}}} \\
\ec

\bc {and} \ec

\bc
{\it \large P. Roy{\footnote {e-mail : pinaki@isical.ac.in}}} \\
\ec

\bc
{\it \large Physics \& Applied Mathematics Unit \\
Indian Statistical Institute \\
Kolkata - 700 108 \\ INDIA } \ec

\vs{.5cm}

\begin{abstract}

We analyze a class of non-Hermitian quadratic Hamiltonians, which
are of the form $ H = {\cal{A}}^{\dagger} {\cal{A}} + \alpha
{\cal{A}} ^2 + \beta {\cal{A}}^{\dagger \ 2} $, where $ \alpha \ ,
\ \beta $ are real constants, with $ \ \alpha \neq \beta $, and
${\cal{A}}^{\dagger}$ and ${\cal{A}}$ are generalized creation and
annihilation operators. Thus these Hamiltonians may be classified
as generalized Swanson models. It is shown that the eigenenergies
are real for a certain range of values of the parameters. A
similarity transformation $\rho$, mapping the non-Hermitian
Hamiltonian $H$ to a Hermitian one $h$, is also obtained. It is
shown that $H$ and $h$ share identical energies. As explicit
examples, the solutions of a couple of models based on the
trigonometric Rosen-Morse I and the hyperbolic Rosen-Morse II type
potentials are obtained. We also study the case when the
non-Hermitian Hamiltonian is ${\cal{PT}}$ symmetric.

\end{abstract}

\vspace{2cm}


\vspace{1cm}


\pagebreak

\section{Introduction}

    The generalization of standard quantum mechanics and quantum field
theory to include complex or non-Hermitian potentials with real
spectrum, has been intensively studied during the last few years
\cite{bender, pt, workshop, applications}, primarily because of
their immense potential for possible applications in a wide range
of phenomena, e.g., nuclear physics \cite{nuclear}, scattering
theory (i.e., complex absorbing potentials) \cite{scattering},
field theory \cite{appl-field}, periodic potentials
\cite{periodic}, quantum cosmology \cite{q-cosmo}, random matrix
theory \cite{rmt}, etc. Initially, the reality of the spectrum was
attributed to the so-called ${\cal{PT}}$ symmetry of the system,
i.e.,
\begin{equation}\label{h-pt}
    H \ \neq H^{\dagger} \qquad ,  \qquad
    H \ {\cal{PT}} ~=~ {\cal{PT}} \ H
\end{equation}
where ${\cal{P}}$ stands for {\it parity}  and ${\cal{T}}$ denotes
{\it time reversal} operators respectively
\begin{equation}\label{pt}
    {\cal{P}}x {\cal{P}} = -x~, \qquad {\cal{P}}p {\cal{P}} = {\cal{T}}
    p {\cal{T}} = -p~, \qquad {\cal{T}}(i. 1) {\cal{T}} = -i. 1
\end{equation}
Such Hamiltonians were found to possess a real and discrete
spectrum when ${\cal{PT}}$ symmetry is exact, i.e., the energy
eigenstates are also the eigenstates of ${\cal{PT}}$ ; \ if not
then ${\cal{PT}}$ symmetry is said to be spontaneously broken and
the energies occur as complex conjugate pairs.

However, it was soon discovered that ${\cal{PT}}$ symmetry is
neither the necessary nor the sufficient criterion for the
spectrum to be real. Subsequent works showed that the necessary
and sufficient condition for a non-Hermitian Hamiltonian to
possess real and discrete spectrum, is its $\eta$-pseudo
Hermiticity, such that $H$ are linear operators acting in a
Hilbert space (generally different from the physical Hilbert
space), and satisfying \cite{mostafa} :
\begin{equation}\label{pseudoH}
    H^{\dagger} \ = \ \eta H \eta ^{-1} \qquad , \qquad
    {\rm{i.e.}} \qquad
    H^{\dagger} \eta \ = \  \eta H
\end{equation}
where $\eta$ is a linear, Hermitian, invertible operator. It may
be mentioned that for a given pseudo-Hermitian operator $H$, the
metric operator $\eta $ is not unique. Furthermore, the
pseudo-Hermiticity of $H$ is equivalent to the presence of an
antilinear symmetry, ${\cal{PT}}$ symmetry being the primary
example \cite{solombrino}. Conversely, a quantum system possessing
an exact antilinear symmetry is pseudo-Hermitian, and is
equivalent to a quantum system described by a Hermitian
Hamiltonian $h$. Thus $H$ may be mapped to $h$, by a similarity
transformation $\rho$ \cite{mostafa,similarity}. For example, let
an eigenvalue (Sturm-Liouville) equation or a differential
operator $H$ act in a complex function space ${\cal{V}}$, endowed
with a positive definite inner product, such that it is described
by the Hilbert space ${\cal{H}}$. In such a case there exists a
mapping from the non-Hermitian $H$ to its Hermitian counterpart
$h$, through a similarity transformation $\rho$ \cite{mostafa2};
i.e.,
\begin{equation}\label{similarity}
    h = \rho H \rho ^{-1}
\end{equation}
with $\rho$ being the unique positive-definite square root of
$\eta$ :
\begin{equation}\label{rho-eta}
    \rho = \sqrt{\eta}
\end{equation}
A relation similar to (\ref{similarity}) holds for observables as
well. For example, if ${\cal{O}}_h$ is an observable in the
Hermitian theory described by $h$, then the corresponding
observable in the pseudo-Hermitian theory is given by
\begin{equation}\label{observable}
    {\cal{O}} = \rho ^{-1} {\cal{O}}_h \ \rho
\end{equation}
Though known for a long time \cite{sudarshan}, the idea of pseudo
Hermiticity was revived after the concept of $\cal{PT}$ symmetry
was introduced a decade ago.

Recently, Swanson analyzed the real but non-Hermitian,
${\cal{PT}}$ symmetric quadratic Hamiltonian \cite{swanson}
\begin{equation}\label{swanson}
    H = \omega a^{\dagger} a + \alpha a^2 + \beta a^{\dagger \ 2}
    \qquad , \qquad \alpha \neq \beta
\end{equation}
where $a^{\dagger}$, $a$ are the Harmonic oscillator creation and
annihilation operators for unit frequency,
\begin{equation}\label{ho-a}
    a = \displaystyle \frac{d}{dx} + x \qquad ,
    \qquad a^{\dagger} = \displaystyle - \frac{d}{dx} + x
\end{equation}
and $\omega, \ \alpha, \ \beta$ are real parameters with
dimensions of inverse time. It was shown that for $\alpha \neq
\beta$, though the Hamiltonian $H$ is non-Hermitian, yet the
eigenvalues were real and positive for $\omega ^2 \geq 4 \alpha
\beta$. This model has attracted the attention of several workers
in recent times, e.g. \cite{musumbu, quesne}. In this work, we
focus our attention on the pseudo-Hermitian generalization of the
Swanson model (\ref{swanson}), which may not necessarily be
${\cal{PT}}$ symmetric. The simplest and most straightforward
generalization would be to consider generalized creation and
annihilation operators $ {\cal{A}} ^{\dagger} $ and ${\cal{A}} $
in place of $a^{\dagger}$ and $ a $, of the form
\begin{equation}\label{a}
    \begin{array}{lcl}
    {\cal{A}} &=& \displaystyle \frac{d}{dx} + W(x) \\ \\
    {\cal{A}} ^{\dagger} &=& \displaystyle - \frac{d}{dx} + W(x)
    \end{array}
\end{equation}
The function $W(x)$, called the pseudo superpotential (in analogy
with conventional supersymmetry), given by
\begin{equation}\label{w}
    W(x) = \displaystyle - \frac{f_0 ^{~ \pr} (x)}{f_0 (x)}
\end{equation}
where $f_0 (x)$ is the ground state wave function of the
Schr\"{o}dinger Hamiltonian {\bf H } $  = {\cal{A}}^{\dagger}
{\cal{A}} $. For the particular case of $W(x)$ being a linear
function in $x$, we get back the Swanson Hamiltonian in
(\ref{swanson}). This is somewhat analogous to the generalization
of the Jaynes-Cummings model to other two level shape-invariant
bound state systems \cite{j-c}, applying the principles of
supersymmetric quantum mechanics \cite{susy}. Thus our starting
Hamiltonian would be
\begin{equation}\label{nonH}
    H = {\cal{A}}^{\dagger} {\cal{A}} + \alpha {\cal{A}} ^2 +
    \beta {\cal{A}}^{\dagger \ 2}  \ , \qquad \alpha \neq \beta
\end{equation}
where $\alpha , \ \beta $ are real, dimensionless constants.
Obviously the model given in (\ref{nonH}) above, is non-Hermitian
for $ \alpha \neq \beta $. In particular, our attempt will be to
give the general formalism for solving such a non-Hermitian
Hamiltonian, and examine the range of values of the parameters for
which the energies are real. This situation is similar to
\cite{swanson}, where real energies were found only when the
parameters satisfied certain constraints. On the other hand since
the Hamiltonian $H$ does not admit real energies for arbitrary
values of the parameters, the model can be termed as conditionally
exactly solvable (CES) \cite{ces}. We shall restrict our study to
$\eta$ pseudo-Hermitian Hamiltonians only, as $\eta$-pseudo
Hermiticity is the necessary and sufficient condition for the
existence of real energies. We shall also find a similarity
transformation $\rho$, mapping the non-Hermitian Hamiltonian $H$
to the Hermitian one $h$, for a certain class of models. It will
be shown that $H$ and $h$ share identical energies. It may be
mentioned here that though the existence of $\eta$, and hence
$\rho$, is guaranteed, it may not always be possible to determine
the Hermitian counterpart $h$ exactly. For example, the
relationship between the non-Hermitian $H$ and its hermitian
entity $h$, was explored in \cite{jones}, for the Swanson model
\cite{swanson} and the $ i g x^3 $ potential. However, in the
first case, $h$ turned out to be a scaled harmonic oscillator,
while in the second model $h$ could be constructed perturbatively
only. It may be mentioned here that the operator method was
employed in \cite{swanson} while we work with the differential
equation directly. The simplicity of the present formalism lies in
the fact that $h$ can be determined in a straightforward manner,
and secondly, $\rho$, and hence $\eta$, can be found exactly, for
the class of non-Hermitian models considered in this work.

The organization of the paper is as follows. In section 2, we
shall give the general formalism for solving a class of
non-Hermitian Swanson model with generalized creation and
annihilation operators. The similarity transformation $\rho$,
between the Hermitian $h$ and the non-Hermitian $H$, is
established in section 3, while the pseudo Hermiticity of $H$ is
shown in section 4. We illustrate our results with the help of a
couple of explicit examples in sections 5 and 6, with Hamiltonians
based on the trigonometric Rosen-Morse I and the hyperbolic
Rosen-Morse II potentials, respectively.  In section 7, a special
sub-class of pseudo-Hermitian Hamiltonians are considered, which
are ${\cal{PT}}$ symmetric as well. Finally, section 8 is kept for
Conclusions and Discussions.

\section{Theory}

As mentioned above, we shall examine a generalization of the
Swanson model , viz., \cite{swanson}
$$
    H =  {\cal{A}}^{\dagger} {\cal{A}} + \alpha {\cal{A}} ^2 +
    \beta {\cal{A}}^{\dagger \ 2}  \ ,  \qquad \alpha \neq \beta
$$
where $\alpha$ and $\beta$ are constants, dimensionless as well as
real. Evidently, $H$ is non-Hermitian for $\alpha \ \neq \ \beta$
for any real $W(x)$. With the help of (\ref{a}), the eigenvalue
equation corresponding to (\ref{nonH}) reads
\begin{equation}\label{schro}
    \begin{array}{lcl}
    H \psi
    &=& \displaystyle \left\{ - \left( 1 - \alpha - \beta
    \right) \frac{d^2}{dx^2}
    + 2 \left( \alpha - \beta \right) W \frac{d}{dx}
    + \left( 1 + \alpha + \beta \right) W^2
    - \left( 1 - \alpha
    + \beta \right) W^{\ \prime}  \right\} \psi \\ \\
    &=& \displaystyle \left\{ - \left( 1 - \alpha - \beta
    \right) \left( \frac{d}{dx} - \frac{\alpha - \beta }{1
    - \alpha - \beta } \ W \right) ^2 +
    \frac{1 - 4 \alpha \beta }{\left( 1 - \alpha - \beta
    \right)} \ W^2
    -  \ W^{\prime} \right\} \psi \\ \\
    &=& E \psi \\ \end{array}
\end{equation}
The term $ \displaystyle \left( - \frac{\alpha - \beta }{1
    - \alpha - \beta } \ W (x) \right)  $ in the parenthesis takes the form of
a complex vector potential and can be eliminated by a gauge
transformation of the form \cite{faria}
\begin{equation}\label{psi}
    \psi (x) = \displaystyle e^{\mu \int W(x) dx } \phi (x) \qquad
    , \qquad \rm{with} \ \ \ \mu = \displaystyle
    \frac{\alpha - \beta} {1 - \alpha - \beta}
    \ , \ \alpha ~+~ \beta ~ \neq ~ 1
\end{equation}
Thus (\ref{schro}) reduces to the well known Schr\"{o}dinger form
\begin{equation}\label{pot}
    h \ \phi (x) ~=~ \displaystyle \left( - \frac{d^2}{dx ^2} \ + \ V(x)
    \right) \phi (x) ~=~ \varepsilon \phi (x)
\end{equation}
where
\begin{equation}\label{phi}
    \begin{array}{lcl}
        V(x) &=& \displaystyle \left( \frac{
        \sqrt{1 - 4 \alpha \beta} }{ 1 - \alpha - \beta } \ W(x)
    \right) ^2  ~-~ \frac{1}{\left( 1 - \alpha -
    \beta \right)} W^{\ \prime} (x) \\ \\
    \ \ \ \ \varepsilon  &=& \displaystyle
    \frac{E}{1 - \alpha - \beta}  \\
    \end{array}
\end{equation}
It is well known from supersymmetric quantum mechanics
\cite{susy}, that $h$ can always be written in a factorizable form
as a product of a pair of linear differential operators $ A \ ,
A^{\dagger} $, as
\begin{equation}\label{factor}
    \begin{array}{lcl}
    h  &=& \displaystyle A^{\dagger} A
    ~+~ \epsilon  \\ \\
    &=& \displaystyle -~ \frac{d^2}{dx^2} ~+~
    {\rm{w}}^2 - {\rm{w}}^{\prime} ~+~ \epsilon
    \end{array}
\end{equation}
where $ \epsilon $ is the factorization energy, and  $ A \ , \
A^{\dagger} $ and ${\rm{w}}(x)$ are given by
\begin{equation}
    A ~=~ \displaystyle \frac{d}{dx} + {\rm{w}}(x) \ , \qquad
    A ^{\dagger} ~=~ \displaystyle - \frac{d}{dx} + {\rm{w}}(x)
    \ , \qquad {\rm{w}} (x) = \displaystyle - \frac{d \ln \varphi _0 (x)}{dx}
\end{equation}
Here $\varphi _0$ is the ground state eigenfunction of
$A^{\dagger} A$ with energy $ \varepsilon_0 $. It may be mentioned
here that SUSY is said to be unbroken when the ground state energy
$ \varepsilon_0 = 0 $.

\vspace{.5cm}

\noindent Evidently, if we can identify the term $V(x)$ in
(\ref{phi}) above, with an exactly solvable potential, then we can
easily find the solutions of $h$. To this end, for further
convenience, $V(x)$ can be identified with a shape-invariant
potential, as using the ideas of supersymmetric quantum mechanics
\cite{susy}, the raising and lowering operator method of harmonic
oscillator can be generalized to a whole class of shape invariant
potentials \cite{sip}, which includes all the analytically
solvable models. To narrow down the class of potentials further,
our strategy would be to write $V(x)$ in (\ref{phi}) in the
supersymmetric form $ {\rm{w}}^2 (x) - {\rm{w}}^{\prime} (x)$ as
given in (\ref{factor}). This identification enables us to find
the energies ($E$) and the eigenfunctions ($\psi$) of the
eigenvalue equation in (\ref{schro}). However, this imposes
certain restrictions on the permissible values of $ \alpha $ and $
\beta $. For real energies, supersymmetric considerations require
the term containing $W^2 (x)$ in the expression for $V(x)$ in
(\ref{phi}), must be positive. Furthermore, $E$ and $\varepsilon$
should have similar behaviour. Hence, the parameters $ \alpha \ ,
\ \beta $ must satisfy the following constraints, irrespective of
the explicit form of $W(x)$ :
\begin{equation}\label{alpha-beta-1}
    \alpha + \beta \ < \ 1 \qquad , \qquad 4 \alpha \beta \ < \ 1
\end{equation}
In addition to the general restrictions imposed on $\alpha \ , \
\beta \ $ in (\ref{alpha-beta-1}), there may be some more
constraints depending on the particular choice of the model,
arising from the normalizability requirement of the wave
functions. We shall illustrate our observations with the help of a
couple of explicit examples in the next section. The fact that
both the models considered here are pseudo-Hermitian will be shown
in a later section.

\vspace{1cm}

\section{Similarity Transformation between $H$ and $h$}

In this section we shall determine a similarity transformation,
mapping the non-Hermitian $H$ to the Hermitian $h$ \cite{mostafa}.
For this purpose we focus our attention on the gauge
transformation $\rho$ relating $\psi (x)$ and $\phi (x)$ in
equation (\ref{psi}); i.e.,
\begin{equation}\label{similar}
\rho = \displaystyle e^{- \mu \int W dx }  \qquad ,
    \qquad \mu = \displaystyle
    \frac{\alpha - \beta} {1 - \alpha - \beta}
\end{equation}
where $ \displaystyle W(x) = - \frac{f _0 ^{\ \pr}(x)}{f_0 (x)} $,
$f_0 (x)$ being the ground state wave function of the
Schr\"{o}dinger Hamiltonian {\bf H} $  = {\cal{A}} ^{\dagger}
{\cal{A}} $. Let $\psi (x)$ be an eigenfunction of $H$ , with
eigenvalue $E$ :
\begin{equation}\label{H-psi}
    H \psi  = E \psi
\end{equation}
Let us now apply the transformation $\rho$ to the above
eigenfunction $\psi (x)$ ; i.e.,
\begin{equation}\label{psi-phi}
    \phi (x) = \rho  \ \psi (x)
\end{equation}
Then (\ref{H-psi}) can be written as
\begin{equation}
    H \rho ^{-1} \phi (x) = E \rho ^{-1} \phi (x) \qquad \qquad {\rm{or}} \qquad
    \qquad \rho \ H \ \rho ^{-1} \ \phi (x) = E \ \phi (x)
\end{equation}
Thus $\phi (x)$ is a solution of the equation $ h \phi  = E \phi $
with the same energy $E$ as in (\ref{H-psi}), provided $H$ is
mapped to $h$ by the similarity transformation in
(\ref{similarity}), viz.,
$$     h = \rho \ H \ \rho ^{-1} $$
As we have observed in this work earlier, $h$ is Hermitian, though
$H$ is non-Hermitian. Thus the similarity transformation $\rho$
given in (\ref{similar}) maps the pseudo-Hermitian Hamiltonian $H$
in the generalized version of the Swanson model to its Hermitian
counterpart $h$. Furthermore, this exact form of the similarity
operator for this class of models, also gives the wavefunctions in
the corresponding Hermitian picture. This will be clarified
further by the explicit models discussed later in this work.

\vspace{1cm}

\section{Pseudo Hermiticity of $H$}

We shall show in this section that, although $H$ in (\ref{schro})
is non ${\cal{PT}}$ symmetric, it is in fact, pseudo-Hermitian,
with respect to a linear, invertible, Hermitian operator $\eta$,
and that it is in fact the square of the similarity transformation
$\rho$, i.e., $ \eta = \rho ^2 $.

We start with the eigenvalue equation $ H \psi = E \psi $, where
$$
    \begin{array}{lcl}
    H &=&  {\cal{A}}^{\dagger} {\cal{A}} ~+~ \alpha {\cal{A}} ^2
    ~+~ \beta {\cal{A}}^{\dagger \ 2}  \\ \\
    &=& \displaystyle - \left( 1 - \alpha - \beta
    \right) \left( \frac{d}{dx} - \frac{\alpha - \beta }{1
    - \alpha - \beta } \ W (x) \right) ^2 +
    \frac{1 - 4 \alpha \beta }{1 - \alpha - \beta} \ W^2 (x)
    -  \ W^{\ \prime} (x)  \\
    \end{array}
$$
Now, let us explore the relationship between $H$ and its adjoint
$H^{\dagger}$, given by
\begin{equation}
    \begin{array}{lcl}
    H^{\dagger} &=&  {\cal{A}}^{\dagger} {\cal{A}}  ~+~
    \alpha {\cal{A}} ^{\dagger \ 2} ~+~ \beta {\cal{A}}^2  \\ \\
    &=& \displaystyle - \left( 1 - \alpha - \beta
    \right) \left( \frac{d}{dx} + \frac{\alpha - \beta }{1
    - \alpha - \beta } \ W (x) \right) ^2 +
    \frac{1 - 4 \alpha \beta }{1 - \alpha - \beta} \ W^2 (x)
    -  \ W^{\ \prime} (x)  \\
    \end{array}
\end{equation}
If we put
\begin{equation}\label{eta}
    \eta = \rho ^2 = \displaystyle e^{- 2 \mu \int W dx }  \ ,
    \qquad \mu = \displaystyle
    \frac{\alpha - \beta} {1 - \alpha - \beta}
\end{equation}
then it can  be shown by straightforward calculations that $H$ and
$H^{\dagger}$ are related by (\ref{pseudoH}), viz.,
$$
    H^{\dagger} \ \eta ~=~ \eta \ H \qquad {\rm{i.e.,}} \qquad
    H^{\dagger} ~=~ \eta \ H \ \eta ^{-1}
$$
In other words, $H$ respects the condition for pseudo Hermiticity
\cite{mostafa}. Thus this approach enables us to determine the
exact form of the pseudo Hermiticity operator $\eta$, which in
turn, is related to the similarity transformation $\rho =
\sqrt{\eta}$.

\vspace{1cm}

\section{A Model based on Trigonometric Rosen-Morse I Potential}

The trigonometric Rosen-Morse I model \cite{susy} is described by
the potential
\begin{equation}\label{rm1}
    V(x) ~=~  \displaystyle  A \left( A - 1 \right) \ \csc ^2  x
    ~+~ 2 B \ \cot  x - A^2 + \frac{B^2}{A^2} \ , \qquad 0
    \leq x \leq \pi
\end{equation}
In the language of supersymmetry, if the potential in (\ref{rm1})
can be written in terms of a superpotential ${\rm{w}}(x)$ as
\begin{equation}
    V(x) = {\rm{w}}^2(x) - {\rm{w}}^{\pr}(x)
\end{equation}
then a suitable ansatz of ${\rm{w}}(x)$ may be given by
\begin{equation}\label{w-ab}
    {\rm{w}}(x) = \displaystyle - A \ \cot  x - \frac{B}{A}
    \qquad , \qquad  A > 0 \ , \ B > 0
\end{equation}
For our model, keeping analogy with the above, we consider the
following form of the function $W(x)$, in the construction of the
generalized annihilation and creation operators ${\cal{A}}$ and
${\cal{A}}^{\dagger}$ in (\ref{a}) :
\begin{equation}\label{wrm1}
    W(x) = \displaystyle - A_1 \ \cot  x - \frac{B_1}{A_1} \qquad
    , \qquad A_1 > 0 \ , \ B_1 > 0
\end{equation}
Obviously, the Hamiltonian in (\ref{schro}) constructed from this
$W(x)$ is non-Hermitian (as well as non ${\cal{PT}}$ symmetric )
for $\alpha \ \neq \ \beta$. Substitution of (\ref{wrm1}) in
(\ref{psi}) yields
\begin{equation}\label{psi-rm1}
    \psi (x) = \displaystyle e^{- \mu _1 x} \sin ^{\mu _2} x \ \
    \phi(x)
\end{equation}
where
\begin{equation}\label{mu12}
    \mu _1 = \displaystyle \frac{B_1}{A_1} \frac{(\alpha - \beta)}{( 1 -
    \alpha - \beta )} \ , \qquad
    \mu _2 = \displaystyle - \frac{A_1(\alpha - \beta)}{( 1 -
    \alpha - \beta )}
\end{equation}
Now, we are interested in real energies only. Additionally, the
wavefunctions must satisfy certain boundary conditions, e.g., well
behaved behaviour at the boundaries $x \ \rightarrow \ 0$ and $x \
\rightarrow \ \pi$, and normalizability requirement. So $ \mu _2 \
> \ 0 $.  These impose further restrictions on $ \alpha $ and $
\beta $, so that they must obey the following condition :
\begin{equation}\label{alpha-beta}
    \alpha \ < \ \beta
\end{equation}
Thus (\ref{pot}) reduces to the trigonometric Rosen Morse I model
in (\ref{rm1}), with potential
\begin{equation}\label{rm11}
    V(x) = \displaystyle
    \sigma \ \csc ^2  x
    ~+~ 2 B_1 \frac{1 - 4 \alpha \beta }{\left( 1 - \alpha - \beta \right) ^2 }
    \ \cot  x
    \left( A_1 ^2 - \frac{B_1 ^2}{A_1 ^2} \right)
    \frac{1 - 4 \alpha \beta }{\left( 1 - \alpha - \beta \right) ^2 }
\end{equation}
where
\begin{equation}\label{sigma1}
    \sigma \ = \ \displaystyle \frac{ A_1 ^2 \left( 1 - 4 \alpha \beta
    \right) - A_1 \left( 1 - \alpha -
    \beta \right)}{\left( 1 - \alpha - \beta \right) ^2}
\end{equation}
so that $A$ and $B$ can be identified with
\begin{equation}\label{ab}
    A \ = \ \displaystyle \frac{1}{2} \pm \frac{\sqrt{1+ 4 \sigma}}{2}
    \qquad , \qquad B \ = \ \displaystyle
    B_1 \frac{1 - 4 \alpha \beta }{ \left( 1 - \alpha -
    \beta \right)^2 }
\end{equation}
Since  $A \ > \ 0 $, only the positive sign is allowed in the
expression for $A$ in (\ref{ab}). Moreover, as is obvious from
(\ref{rm11}), for the existence of bound states, $ \sigma \ > 0 $.
Since $ A_1 \neq 0 $, hence this condition requires
\begin{equation}\label{a-1}
    \displaystyle A_1 \ > \ \frac{1 - \alpha - \beta }
    {1 - 4 \alpha \beta}
\end{equation}
The energy eigenvalues and the corresponding eigenfunctions of
(\ref{rm1}) are well known \cite{susy}
\begin{equation}\label{rm1-e}
    \epsilon _n ~=~
    \displaystyle \left( A+n \right) ^2 - \frac{B^2}{\left(
    A+n \right)^2} ~-~ A^2 ~+~ \frac{B^2}{A ^2} \qquad ,
    \qquad \qquad  n=0, 1, 2, \cdots
\end{equation}
Therefore (\ref{rm11}) has solutions
\begin{equation}\label{rm11-e}
    \varepsilon _n ~=~
    \displaystyle \left( A+n \right) ^2 ~-~ \frac{B^2}{\left(
    A+n \right)^2} ~-~
    \left( A_1 ^2 - \frac{B_1 ^2}{A_1 ^2} \right)
    \frac{1 - 4 \alpha \beta }{\left( 1 - \alpha - \beta \right) ^2 }
\end{equation}
where $A$ and $B$ are given in terms of $A_1$ and $B_1$ through
(\ref{ab}), and the wavefunctions are
\begin{equation}\label{rm11-phi}
    \phi_n (x) ~\approx~ \displaystyle \left( y^2 - 1 \right)
    ^{- \frac{(A+n)}{2}}
    \displaystyle e ^{\left( \frac{B}{A+n} \right) x} \
    P_n ^{\left( s_+ \ , \ s_-  \right) }
    (y) \ \ , \ \ \ y = i \ \cot x
\end{equation}
\begin{equation}\label{s}
    \displaystyle s_{\pm} ~=~ -A - n  \pm i \frac{B}{(A+n)}
\end{equation}
In (\ref{rm11-phi}) above, $ P_n ^{\left( s_+ \ , \ s_-  \right) }
(y) $ are the standard Jacobi polynomials \cite{handbook}. Using
(\ref{rm11-e}) and (\ref{rm11-phi}) one can easily obtain the
energies and eigenfunctions of the eigenvalue equation in
(\ref{schro}), for this particular model as :
\begin{equation}\label{e}
    E_n = \displaystyle \left( 1 - \alpha - \beta \right)
    \varepsilon_n
\end{equation}
\begin{equation}\label{psi1}
    \psi_n (x) ~\approx~ \displaystyle
    e ^{\left\{ \frac{B}{(A+n)} - \mu_1 \right\} x}
    \sin ^ {A + n + \mu _2}  x  \
    P_n ^{\left( s_+ \ , \ s_- \right) }
    (y) \ \ , \ \ \ y = i \ \cot x
\end{equation}
Thus one gets the complete solution of the non-Hermitian
Hamiltonian in (\ref{nonH}), by reducing it to the corresponding
Hermitian system.

\vspace{.5cm}

\noindent{\large \bf Choice of parameters}

\vspace{.25cm}

To show that the solutions (\ref{e}) and (\ref{psi1}) actually
exist, it is necessary to show that there are parameter values
actually satisfying (\ref{alpha-beta-1}), (\ref{alpha-beta}) and
(\ref{a-1}). There may be innumerable such combinations of
$\alpha$ , $\beta$, $A_1$ and $B_1$. We show a few possible values
of these parameters in Table 1. In each case, the potential is
given as in (\ref{rm11}), with solutions $\psi _n (x) $ given in
(\ref{psi1}) above, and energies in (\ref{e}).

\vspace{1cm}

\noindent {\bf Table 1 : } Some values of the parameters for the
model with $W(x)$ as given in (\ref{wrm1})

\vspace{.3cm}

\begin{tabular}{|c|c|c|c|c|c|c|c|c|c|c|c|c|}
  \hline  $ \displaystyle $
  $\alpha$ & $\beta$ & $\alpha + \beta $ &
  $4 \alpha \beta$  & $A_1$ & $B_1$ & $ \mu _1$ & $\mu _2$ &
  $\sigma$ & $A$ & $B$ & $E_n$  \\
  \hline
  1/4 & 1/2  & 3/4 & 1/2 & 3/2 & 1/8 & - 1/12 & 3/2 &
  12 & 4 & 1 &
  $ \displaystyle \frac{1}{4} \varepsilon _n $  \\ \hline
  1/4 & 2/3 & 11/12 & 2/3 & 1 & 1/2 &
  -5/2 & 5 & 36 & 6.52  & 24 &
  $ \displaystyle \frac{1}{12} \varepsilon _n $ \\ \hline
  1/8 & 3/4 & 7/8 & 3/8 & 1 & 2 &
  -10 & 5 & 32 & 6.18 & 80 &
  $ \displaystyle \frac{1}{8} \varepsilon _n $ \\ \hline
  1/3 & 1/2 & 5/6 & 2/3 & 1 & 2 &
  -2 & 1 & 6 & 3 & 36 &
  $ \displaystyle \frac{1}{6} \varepsilon _n $ \\ \hline
\end{tabular}

\vspace{.5cm}

One can check the nature of the non Hermitian Hamiltonian and the
corresponding Hermitian equivalent for this model. For example,
for the values of parameters in the first line of Table 1, the
starting non-Hermitian equation (\ref{schro}), is given by
\begin{equation}\label{nH-trigo}
\begin{array}{lcl}
    H \psi (x) &=& \displaystyle  \left\{- \frac{1}{4}
    \frac{d^2}{dx^2} ~+~ \left(
    \frac{18  \ \cot \ x + 1}{24} \right) \frac{d}{dx}
    \displaystyle +~ \frac{33}{16} \left( \csc x \right) ^2
    + \frac{7}{16} \cot x - \frac{2261}{ 576}  \right\} \psi (x) \\ \\
    &=& E \psi(x)
    \end{array}
\end{equation}
With the help of the similarity transformation in (\ref{similar}),
the above non-Hermitian equation is transformed to the Hermitian
one
\begin{equation}\label{h-trigo}
    h \phi (x) = \left\{ - \frac{d^2}{dx^2} +
    12 \ \csc ^2 x ~+~ 2 \ \cot x -
    \frac{323}{18} \right\} \phi (x)
    = \varepsilon \phi(x)
\end{equation}
where $E = \displaystyle \frac{1}{4} \varepsilon  $, and $ \psi $
and $ \phi $ are related by
\begin{equation}
    \psi (x) = \displaystyle e^{\frac{1}{12} x}
    \sin ^{\frac{3}{2} x} \phi(x)
\end{equation}
Since equation (\ref{h-trigo}) can be solved exactly, one can use
its solutions to find the energies and eigenfunctions of the
non-Hermitian equation in (\ref{nH-trigo}).

It is worth mentioning here that a second order linear
differential equation can have only two linearly independent
solutions. For the model discussed in this section, only one of
the solutions is normalizable in the Hermitian picture. So the
second solution is not considered. It can be checked by
straightforward algebra that even when they are mapped to the non
Hermitian picture, the second solution does not have well defined
behaviour at the boundaries, irrespective of the fact whether the
parameters $\alpha \ , \ \beta $ obey the constraints
(\ref{alpha-beta-1}) or not. Furthermore, for the acceptable set
of solutions in the Hermitian picture, well-defined behaviour of
the eigenfunctions at the boundaries, and the normalization
condition, hold only when the parameters $\alpha \ , \ \beta$ ,
etc. satisfy the constraints (\ref{alpha-beta-1}), (\ref{w-ab}),
(\ref{alpha-beta}) and (\ref{a-1}). Detailed but simple
calculations reveal that the constraints remain unaltered when one
moves from the Hermitian to the non Hermitian picture. Hence the
solutions given here represent the complete set, in both the
Hermitian as well as the non Hermitian picture.

\vspace{1cm}

\section{A Model based on Hyperbolic Rosen-Morse II Potential}

As a second non-Hermitian as well as non ${\cal{PT}}$ symmetric
example, we shall consider a model based on the hyperbolic
Rosen-Morse II potential, given by \cite{susy}
\begin{equation}\label{rm2}
\begin{array}{lcl}
    V(x) &=&  \displaystyle - a \left( a + 1 \right) \ {\rm{sech}} ^2 x
    ~+~ 2 \ b \ \tanh x ~+~ a^2 ~+~ \frac{b^2}{a^2} \ , \ \ \ b
    < a^2  , \ - \infty \leq x \leq \infty \\ \\
    &=& {\rm{w}}^2 (x) ~-~ {\rm{w}} ^{\prime} (x)
    \end{array}
\end{equation}
with the superpotential ${\rm{w}}(x)$ of the form
\begin{equation}
    {\rm{w}}(x) = \displaystyle a \ \tanh x + \frac{b}{a} \
    , \qquad \qquad  b < a^2 \ \ \ \rm{and} \ \ \ a \ , \ b \ > \ 0
\end{equation}
Analogous to the previous example, to construct the generalized
annihilation and creation operators in (\ref{a}), we take the
following ansatz for $W(x)$ :
\begin{equation}\label{wrm2}
    W(x) = \displaystyle A_2 \ \tanh x + \frac{B_2}{A_2} \
    , \qquad \qquad B_2 < A_2 ^2 \ \ \ \rm{and} \ \
    \ A_2 \ , \ B_2 \ > \ 0
\end{equation}
Proceeding along the lines similar to the earlier example, the
eigenvalue equation in (\ref{pot}) reduces to that of the
well-known hyperbolic Rosen Morse-II model in (\ref{rm2}), with
the potential
\begin{equation}\label{hrm2}
    V(x) ~=~ \displaystyle - \chi \ {\rm{sech}} ^2 \ x
    ~+~ 2 B_2 \frac{1 - 4 \alpha \beta }{\left( 1 - \alpha - \beta \right) ^2 }
    \ \tanh x  + \left( A_2 ^2 + \frac{B_2 ^2}{A_2 ^2} \right)
    \frac{1 - 4 \alpha \beta }{\left( 1 - \alpha - \beta \right) ^2 }
\end{equation}
provided one makes the identification
\begin{equation}\label{aabb}
    a = \displaystyle - \frac{1}{2} \pm \frac{\sqrt{1+ 4 \chi}}{2}
    \ , \qquad \qquad
    b = \displaystyle B_2 \frac{1 - 4 \alpha \beta }{ \left( 1 - \alpha -
    \beta \right)^2 }
\end{equation}
with
\begin{equation}\label{chi}
    \chi = \displaystyle \frac{ A_2 ^2 \left( 1 - 4 \alpha \beta
    \right) + A_2 \left( 1 - \alpha -
    \beta \right)}{\left( 1 - \alpha - \beta \right) ^2}
\end{equation}
Once again, since $a \ > \ 0 $, only the positive sign is allowed
in (\ref{aabb}) in the expression for $a$.  Thus the solutions
$\phi (x)$ of the eigenvalue equation in (\ref{pot}) with the
potential in (\ref{hrm2}), are related to the solutions $\psi(x)$
of $H$ in (\ref{schro}) by the substitution in (\ref{psi}) :
\begin{equation}\label{psi-rm2}
    \psi(x) = \displaystyle e^{\mu _1 x} \ \cosh ^{\mu _2} x \
    \phi(x)
\end{equation}
with
\begin{equation}
    \mu _1 = \displaystyle \frac{B_2}{A_2} \frac{(\alpha - \beta)}{( 1 -
    \alpha - \beta )} \ , \qquad
    \mu _2 = \displaystyle  \frac{A_2(\alpha - \beta)}{( 1 -
    \alpha - \beta )} \ , \qquad \alpha + \beta \ < \  1
\end{equation}
For the eigenfunction to be well behaved at $x \ = \ \pm \ \infty
$, $\mu_2 $ should be negative, so that $\alpha \ < \  \beta $.
Additionally, $\displaystyle  |\mu _2|  >  |\mu _1|$, which, in
turn, requires $ B_2  <  A_2 ^2$, as already mentioned in
(\ref{wrm2}). These constraints on $\alpha \ , \ \beta$, which
depend on the explicit form of the model considered, are in
addition to the ones in (\ref{alpha-beta-1}). The energy
eigenvalues and eigenfunctions to (\ref{hrm2}) are respectively
given by,
\begin{equation}
    \varepsilon _n = \displaystyle - (a-n)^2 - \frac{b^2}{(a-n)^2}
    + \displaystyle
    \left( A_2 ^2 + \frac{B_2 ^2}{A_2 ^2} \right)
    \frac{1 - 4 \alpha \beta }{\left( 1 - \alpha - \beta \right) ^2 }
    \qquad , \qquad n < a
\end{equation}
\begin{equation}
    \phi _n (x) \approx \displaystyle (1-y)^{s_+/2} \ (1+y)^{s_-/2}
    P^{(s_+ \ , \ s_-)} _n (y) \ , \qquad y = \tanh x
\end{equation}
where
\begin{equation}\label{spm}
    s_{\pm} ~=~ \displaystyle a - n \pm \frac{b}{a-n}
\end{equation}
$ P^{(s_+ \ , \ s_-)} _n (y) $ are the Jacobi polynomials
\cite{handbook}, and $a , \ b$ are given in terms of $A_2 , \ B_2$
through (\ref{aabb}) and (\ref{chi}). The corresponding energies
and the eigenfunctions of the eigenvalue equation in (\ref{schro})
are obtained as :
\begin{equation}\label{e-2}
    E_n = \displaystyle \left( 1 - \alpha - \beta \right)
    \varepsilon_n \qquad , \qquad n = 0, 1, 2, \cdots < a
\end{equation}
\begin{equation}\label{psi2}
    \psi _n (x) \approx \displaystyle (1-y)^{\left( s_+ - \mu _2 \right)/2}
    \ (1+y)^{\left(s_- - \mu _2 \right)/2}
    e^{\mu _1 x} \ P^{(s_+ \ , \ s_-)} _n (y) \ , \qquad y = \tanh x
\end{equation}
For normalizable functions with real energies, and well defined
behaviour at $ x \ \rightarrow \ \pm \infty $, the constraints
given in (\ref{alpha-beta}) hold here, too.

\vspace{.5cm}

\noindent{\large \bf Choice of parameters}

\vspace{.35cm}

Analogous to the previous case, here, too, the solutions
(\ref{e-2}) and (\ref{psi2}) are acceptable in certain ranges of
the parameters $\alpha, \ \beta$, satisfying (\ref{alpha-beta-1})
and (\ref{alpha-beta}). Many such combinations are possible. We
list a few cases in Table 2.


\pagebreak

\noindent {\bf Table 2 : } Some values of the parameters for the
model with $W(x)$ as given in (\ref{wrm2})

\vspace{.3cm}

\begin{tabular}{|c|c|c|c|c|c|c|c|c|c|c|c|c|}
  \hline  $ \displaystyle $
  $\alpha$ & $\beta$ & $\alpha + \beta $ & $4 \alpha \beta$
  & $A_2$ & $B_2$ & $ \mu _1$ & $\mu _2$ &
  $\chi$ & $a$ & $b$ & $E_n$  \\
  \hline
  1/4 & 1/2 & 3/4 & 1/2 & 3/2 & 1/4 & -1/6 & -3/2 &
  24 & 4.42  & 2 &
  $ \displaystyle \frac{1}{4} \varepsilon _n $ \\ \hline
  1/3 & 1/2 & 5/6 & 2/3 & 1 & 1/8 & - 1/8 & -1 & 18 &
  3.74 & 3/2 & $ \displaystyle \frac{1}{6} \varepsilon _n $
  \\ \hline
  1/6 & 1/3 & 1/2 & 2/9 & 3/2 & 1/2 & -1/9 & -1/2 &
  10 & 2.70 & 1/2 &
  $ \displaystyle \frac{1}{6} \varepsilon _n $  \\ \hline
  1/3 & 1/2  & 5/6 & 2/3 & 1/2 & 1/8 & - 1/4 & -1/2 &
  6 & 2 & 4 &
  $ \displaystyle \frac{1}{6} \varepsilon _n $  \\ \hline

\end{tabular}

\vspace{.5cm}

The discussion at the end of Section 5, on the completeness of
solutions, holds for this model as well.

\vspace{1cm}

\section{${\cal{PT}}$ invariant Generalized Swanson Model }

 The importance of quantum systems with ${\cal{PT}}$
symmetry has already been discussed briefly earlier in this work.
So in this section we consider a particular case of the
non-Hermitian Hamiltonian in (\ref{nonH}) which is symmetric under
the combined effect of ${\cal{PT}}$. For $H$ to be invariant under
${\cal{PT}}$ symmetry, ${\cal{A}}$ and ${\cal{A}}^{\dagger}$
should also be ${\cal{PT}}$ invariant. For this purpose, following
the ${\cal{PT}} $ transformations in (\ref{pt}), the operators
${\cal{A}}$ and ${\cal{A}}^{\dagger}$ should transform under
parity and time reversal as
\begin{equation}\label{a-pt}
    {\cal{P}} \ : \ {\cal{A}} \ ({\cal{A}}^{\dagger}) \
    \rightarrow \ - \  {\cal{A}} \ ({\cal{A}}^{\dagger})
    \qquad \qquad , \qquad \qquad
    {\cal{T}} \ : \ {\cal{A}} \ ({\cal{A}}^{\dagger}) \
    \rightarrow \ {\cal{A}} \ ({\cal{A}}^{\dagger})
\end{equation}
This is possible only if $W(x)$ transforms under ${\cal{PT}}$ as
\begin{equation}\label{w-pt}
    ({\cal{PT}}) \ W(x) \ ({\cal{PT}})^{-1} = - \ W(x)
\end{equation}
Incidentally, the pseudo superpotentials considered in
(\ref{wrm1}) and (\ref{wrm2}) fail to obey the above condition
(\ref{w-pt}) for non zero $B_1$ or $B_2$.

\subsection{Model based on Trigonometric Rosen Morse
potential with $B _1 = 0$ }

If we consider the particular case $B _1 = 0$ in the trigonometric
Rosen Morse model,
\begin{equation}
    W(x) = - A_1 \cot x \qquad , \qquad A_1 > 0
\end{equation}
then, the pseudo superpotential satisfies the condition
(\ref{w-pt}), and the model, in addition to being $\eta$
pseudo-Hermitian, is also ${\cal{PT}}$ symmetric. In such a case,
both $B_1$ and $\mu_1$ are zero. Thus, though the constraints on
$\alpha \ , \beta$ remain unaltered, the columns $B$, $B_1$ and
$\mu _1$ are absent in Table 1. For the parameter values already
discussed above, the potential in (\ref{rm11}) assumes the simple
form
\begin{equation}
    V(x) ~=~ \displaystyle A(A+1) \ \csc ^2 x - A^2
\end{equation}
with energies
\begin{equation}
    \varepsilon_n = (A+n)^2 - A^2
\end{equation}
Thus the solutions of the eigenvalue equation in (\ref{schro}) are
explicitly given by
\begin{equation}\label{psi1-cal}
    \psi_n (x) ~\approx~ \displaystyle
    \left( \sin  x \right) ^ {A + n + \mu _2} \
    P_n ^{\left( -A - n , -A - n
    \right) }
    (i \ \cot x)
\end{equation}
with energies $ E_n = \displaystyle \frac{1}{1-\alpha - \beta} \
\varepsilon_n $.

\subsection{Model based on Hyperbolic Rosen Morse
potential with $B _2 = 0$ }

Analogous to the previous model, for the particular case $B_2 =
0$, both $\mu _1 $ and $b$ turn out to be zero, and this
non-Hermitian model, too, becomes ${\cal{PT}}$ symmetric. The
potential in (\ref{hrm2}) reduces to
\begin{equation}
    V(x) = - a (a+1) sech ^2 \  x + a^2
\end{equation}
having real energies
\begin{equation}
    \varepsilon _n = \displaystyle \left\{ - (a-n)^2 + a^2
    \right\} \qquad , \qquad n = 0, 1, \cdots , < a
\end{equation}
and solutions
\begin{equation}
    \psi _n (x) \approx \displaystyle (sech \ x )^{(s - \mu_2)}
     \ P^{(s \ , \ s)} _n (\tanh x)
\end{equation}
where
\begin{equation}\label{s-pm}
    s_+ = s_- = s =  a-n
\end{equation}
This enables us to find the eigenfunctions and eigenvalues of the
original equation in (\ref{schro}). Once again, the restrictions
on $\alpha \ , \ \beta$ are the same as before, viz., conditions
(\ref{alpha-beta-1}) and (\ref{alpha-beta}), but the columns under
$b \ , \ B_2 \ , \ \mu_1 $ are missing from Table 2.

\vspace{1cm}

\section{Conclusions :}

To conclude, we have studied a class of pseudo-Hermitian
Hamiltonians (not necessarily ${\cal{PT}}$ symmetric) of the form
$ H = {\cal{A}}^{\dagger} \ {\cal{A}} ~+~ \alpha \ {\cal{A}} ^2
~+~ \beta \ {\cal{A}}^{\dagger \ 2}  \ $, where $\alpha$ and
$\beta$ are real, dimensionless constants ($ \alpha \neq \beta $),
and ${\cal{A}}^{\dagger} $ and ${\cal{A}}$ are generalized
creation and annihilation operators. Incidentally, Swanson studied
a similar model \cite{swanson}, although with harmonic oscillator
creation and annihilation operators only. Two explicit examples
are considered in this work --- viz., models based on the
trigonometric Rosen-Morse I and the hyperbolic Rosen-Morse II type
potentials. It is observed that the eigen energies are real for a
certain range of values of the parameters $\alpha, \beta$. A
similarity transformation $\rho$, mapping the non-Hermitian
Hamiltonian $H$ to a Hermitian one $h$, is also obtained. It is
observed that $H$ and $h$ share identical energies. Furthermore,
the linear operator $H$ is pseudo-Hermitian with respect to the
square of the similarity transformation $ \eta = \rho ^2 $.  This
straightforward approach provides us a simple way of determining
the similarity transformation $\rho$,  the metric operator $\eta$,
as well as the corresponding Hermitian Hamiltonian $h$.

As a mathematical interest, one can also start with the
pseudo-Hermitian (but non ${\cal{PT}}$ symmetric) model $H_1$,
given by $ \displaystyle
    H_1 =  {\cal{A}} \ {\cal{A}} ^{\dagger} ~+~ \alpha {\cal{A}} ^2
    ~+~ \beta {\cal{A}} ^{\dagger \ 2}  $,
and proceed as shown in this work. This is possible because of the
fact that while in the case of the Swanson model $ [ a ,
a^{\dagger} ] =  constant $,  the commutator of the generalized
annihilation and creation operators ${\cal{A}}$ and
${\cal{A}}^{\dagger}$ is quite non trivial : $ [ {\cal{A}} ,
{\cal{A}}^{\dagger} ] =  2 W ^{\ \prime} (x) $.

It would be interesting to repeat this analysis with non-Hermitian
complex potentials. As an example, one may write equation
(\ref{wrm2}) as $ \displaystyle W(x) =  A_2 \tanh x + i
\frac{B_2}{A_2} $. $H$ obtained in this way, is non-Hermitian,
complex and ${\cal{PT}}$ symmetric, and the procedure is valid for
such a case as well. Another interesting area of study would be to
examine the applicability of this procedure to non-shape invariant
exactly solvable potentials, including QES (quasi-exactly
solvable) and CES (conditionally exactly solvable) ones.

\vspace{1cm}

\section{Acknowledgment}

This work was partly supported  by SERC, DST, Govt. of India,
through the Fast Track Scheme for Young Scientists (DO No. SR /
FTP / PS-07 / 2004), to one of the authors (AS). The authors also
thank the referees for their valuable comments.

\end{document}